\documentclass[aip,jcp,reprint]{revtex4-1} 
\usepackage{verbatim}
\usepackage{amsmath}
\usepackage{tikz}
\usetikzlibrary{matrix,chains,positioning,decorations.pathreplacing,arrows}
\usepackage{float}
\usepackage{graphicx}
\usepackage{amsmath}    
\usepackage{graphicx}   
\usepackage{dcolumn}
\usepackage{color}
\usepackage{ulem}
\usepackage{soul}
\usepackage{caption}
\usepackage{subcaption}
\usepackage{algorithm} 
\usepackage{array}
\usepackage{booktabs}
\usepackage{multirow}

\usepackage{booktabs}

\newcommand*{\Comb}[2]{{}^{#1}C_{#2}}

\usepackage[colorlinks=true,linkcolor=blue,anchorcolor=black,citecolor=blue,filecolor=black,menucolor=black,runcolor=black,urlcolor=blue]{hyperref}

\begin{document}

\title{ A Deep Autoencoder Framework for Discovery of Metastable Ensembles in Biomacromolecules }

\author{Satyabrata Bandyopadhyay}\affiliation{Tata Institute of Fundamental Research, Center for Interdisciplinary sciences, Hyderabad 500046, India}\author{Jagannath Mondal} \email{jmondal@tifrh.res.in, +914020203091} \affiliation{Tata Institute of Fundamental Research, Center for Interdisciplinary sciences, Hyderabad 500046, India} 
\date{\today} 

\begin{abstract}

   Mini-proteins and peptides manifest dynamic conformational fluctuation and involve mutual interconversion among metastable states. A robust mapping of the conformational landscape underlying mini-proteins and peptides often requires low-dimensional projection of the conformational ensemble along optimized collective variables. However, the traditional choice for the collective variable (CV) is often limited by user-intuition and prior knowledge about the system, which lacks a rigorous assessment of their optimality over other candidate CVs. To address this issue, we propose a generic approach in which we first choose the possible combinations of inter-residue C$\alpha$-distances within a given macromolecule as a set of input CVs. Subsequently we derive a non-linear combination of latent-space embedded collective variables via auto-encoding the unbiased MD simulation trajectories within the framework of feed-forward neural network. We demonstrate the ability of the derived latent space variables in elucidating the conformational landscape in three hierarchically complex systems. When the conformational dynamics is resolved along the latent space CVs, it identifies key metastable states of a bead-in-a-spring polymer. The combination of the adopted dimensionally reduction technique with a Markov state model, built on the derived latent space, efficiently projects the free energy landscape of GB1 $\beta$-hairpin, revealing multiple spatially well-resolved and kinetically well-separated metastable conformations. A quantitative comparison based on variational approach to Markov Process of the auto encoder-derived latent-space CVs with the ones obtained from independent component analysis (PCA or TICA) confirms the optimality of the former. Finally, as a practical application, we demonstrate that the auto-encoder derived CVs successfully predict the reinforced folding of Trp-cage mini-protein in an aqueous osmolyte solution.       

  \end{abstract}

\maketitle
\newpage

\section{Introduction}

      Bio-macromolecules are intrinsically complex and are often associated with rugged thermodynamic and kinetic conformational landscapes\cite{funnel_pathways_Bryngelson_1995,rugged_fes_Dill_1997}. This is more prominent in case of small proteins and peptides due to their large conformational fluctuations, which give rise to dynamically interconverting metastable ensemble of conformations. However, the large number of degrees of freedom associated with the biomolecules eludes spatial and temporal dissection of underlying conformational landscape. A popular approach to circumvent this issue is to reduce the dimensionality of the system. This approach has motivated  derivation of appropriate collective variables (CV) or feature-space for the system of interest. Generally these CVs are often used to project the conformational ensembles of biomacromolecules along these subs-spaces.  The quality of chosen CVs is crucial for suitable projection and identification of key macrostates. Hence a rigorous assessment and optimization of CVs would play very important roles in elucidating the complicated biomolecular dynamics.       
        
        The process of hunting the right CVs for distinguishing the ensembles of biomacromolecular conformations is a non-trivial task. Traditionally, the choice has been mostly driven as well as limited by the user's input and personal experience on the system of interest. This process often forces one either to choose certain pre-decided CVs (mostly single or a pair of CVs) or requires arbitrary attempts via trial or error. The projection of conformational motions along those user-defined CVs does not guarantee exhaustive identification of all metastable conformations. Moreover, this approach most often runs into the risk of missing out on important conformational basin in those projected basins. Significantly, the process lacks a rigorous assessment of the chosen CVs for their optimality. This is particularly relevant for enhanced sampling techniques namely umbrella sampling\cite{Umbrella_sampling}  or, metadynamics\cite{Metadynamics_Laio,Metadynamics_review} simulation which accelerate the exploration of phase space via biasing certain CVs, in which inappropriate choice of CVs can result in misleading free energetics.
        
        \begin{figure}[H]
        	\includegraphics[clip,width=1.0\columnwidth,keepaspectratio]{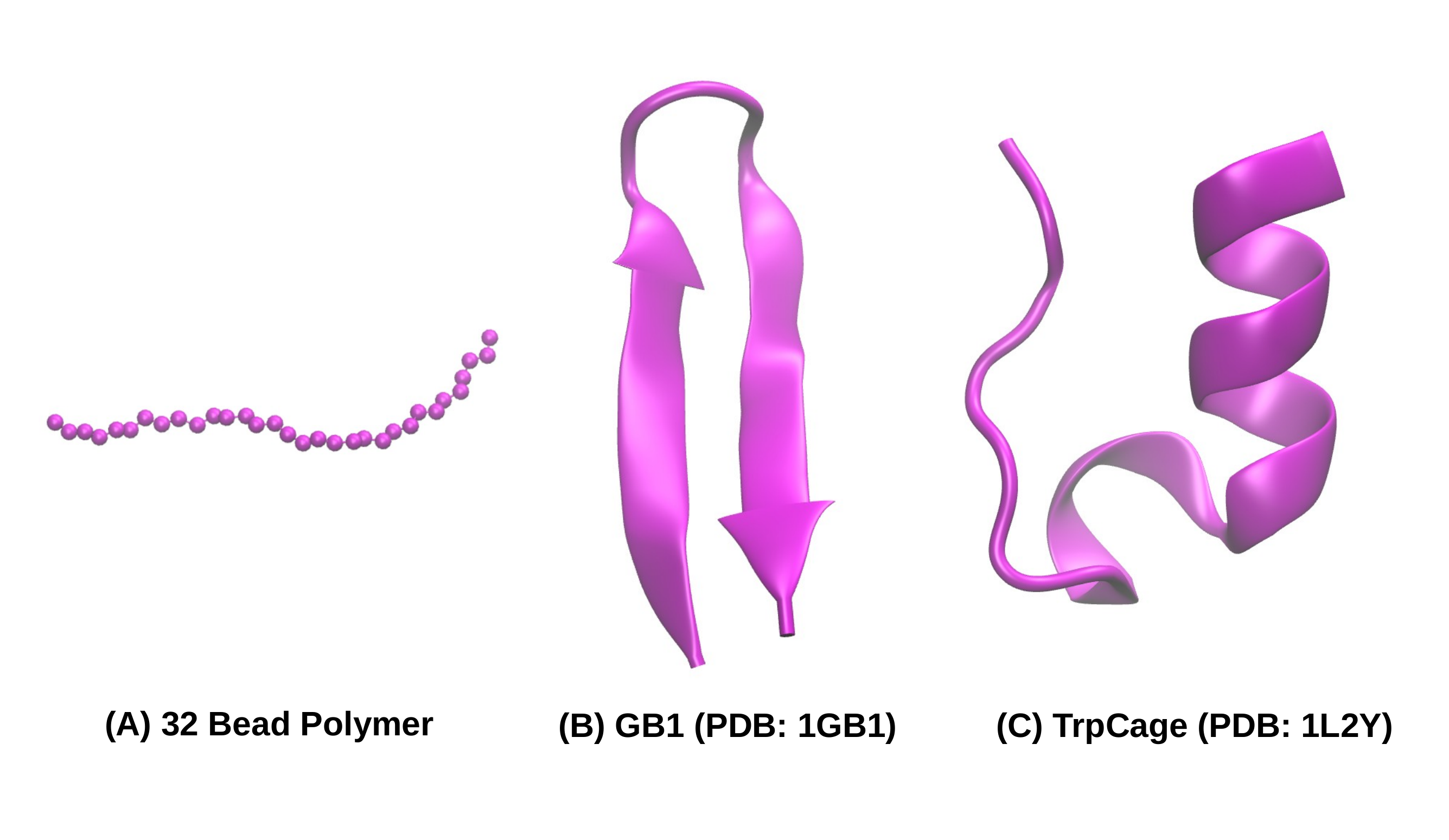}
        	\caption{Systems used in the current work. }
        	\label{snap}
        \end{figure}
        
       A continued thrust over last decade has been focussed on the invention of new approaches for assessment and optimization of the candidate CVs for a biomolecular process. This has resulted in emergent directions involving maximum caliber based approach like Spectral gap optimization of order parameter\cite{Tiwary_2016_sgoop}. Independent component analysis (ICA) based approaches have also received prominence for their ability to efficiently combine multiple CVs in a linear fashion. In particular, principal component analysis (PCA)\cite{PCA_ref1} has become a regular approach for projecting the conformational space along the direction of maximum variance. Lately, time-structured independent component analysis (TICA) \cite{TICA_ref1,TICA_ref2} is gaining traction for its efficiency in projecting the conformational landscape along kinetically slow directions. The implementation of TICA has seen successful application in identifying metastable states and quantitatively assessing the contribution of constituent CVs in the resultant CVs. TICA has now been well-integrated as a key method of choice as a dimension-reduction technique for building Markov state Models for kinetics. 
        
        However, the inherently linear nature of ICA-based CVs poses a key restriction in CV optimization process.  The realization of limitation in linearly combined CVs has paved the way for more general machine learning based approaches which are inherently nonlinear in nature. Machine learning provides a systematic way-out to discover data-driven features of CVs in Molecular Dynamics (MD) simulations. Conceptually, these approaches posit that the MD in the large-dimensional (3N-dimensional) space of the Cartesian coordinates of macromolecule(s) are effectively encoded in a low-dimensional `intrinsic
manifold' containing the slow dynamics to which the remaining fast degrees of freedom are
filtered out. The validity of this assumption has been demonstrated in multiple investigations and in all cases, the approach has aimed at obtaining a small number of collective modes from a large set of coupled degrees of freedoms. In this regard, developments including diffusion maps\cite{Ferguson_2011_diffusion_map} and sketch maps\cite{Ardevol_2015_sketch_map}  have demonstrated great success in parameterising nonlinear intrinsic manifolds of complex macromolecules by discovering nonlinear CVs for capturing conformational changes. 

            In the current work, we harness the efficiency of a popular deep-learning algorithm, namely auto associative encoder, for an efficient derivation of optimized CVs for a set of conformationally dynamic macromolecules. The recent times have seen successful implementation of auto encoders and its variants for CV discovery and predicting the conformational free energetics and dynamics.  The purpose of dimensional reduction followed by feature extraction has become an active area of research\cite{Hinton_2006_science}. The reduced dimensions obtained by artificial neural network holds promises of becoming powerful collective variables hence would fulfil the purpose of collective variable discovery\cite{Chen_2018_cv_discovery_in_enhanced_sampling, Wehmeyer_2018_Noe_TAE, Ribeiro_2018_rave, Hern_ndez_2018_variation_encoding,Sultan_2018_Transferable_NN,Lemke_2019_Encoder_Map,Chen_2019_time_lagged_encoder}.  Wang and Ferguson\cite{Wang_2018_Manifold_Learning} has applied machine learning to generate the folding funnel using delay embeddings on time series data and manifold learning. Tiwary et al.\cite{Wang_2020_Tiwary_ML_enhancing_md} have applied the machine learning techniques to analyze and enhance the molecular dynamics simulations. 
            
            On a similar spirit, this article  adopts a feed-forward neural network based framework. A key advantage of this technique is that one can choose extensively large set of CVs as input and can reduce this into a complex low-dimensional embedding in the `latent space', provided that the input is autoencoded in the output. Due to its ability to handle large set of data as an input, the choice of input CVs does not have any restriction in its number.  Accordingly, in the current work, we go beyond the conventional and user-intuitive choice of CVs (Rg, RMSD, number of contacts etc) for biomolecules. In stead, we use a large set of inter-residue distances  as the input CVs and auto-encode the choice within the framework of feedforward neural networks. The approach gives rise to a minimal number of latent-space CVs in which the original input CVs are compressed. We first show that the latent space CVs, resulting from this approach, can efficiently identify the key intermediates in a collapse dynamics of bead-in-a-spring polymer in solvent. Second, the projection of MD simulation trajectories of GB1 $\beta$-haripin along the latent space leads to identification of distinct metastable states. A quantitative comparison, based on variational approach to Markov processes (VAMP), confirms the superiority  of derived latent space CVs over PCA and TICA derived CVs. A Markov state model developed on the latent space provides robust spatial and temporal resolution of conformational landscape of GB1 $\beta$-hairpin. Finally, the technique is successfully assessed for its ability to capture the osmolyte-induced folding of Trp-cage mini-protein.

\section{Material and Methods}
\textit{Description of Systems:}

      Three systems of increasing hierarchical complexity are used in the work: 1) A model bead-in-a-spring polymer, 2) 16-residue GB1 $\beta$-hairpin Ace-$^{41}$GEWTYDDATKTFTVTE$^{56}$-NMe, a polypeptide corresponding to C-terminal domain of GB1 protein. 3) Trp-cage, a 20-residue mini-protein (see Figure ~\ref{snap}). A large set of all-atom Molecular Dynamics simulation trajectories for each of the systems act as the input source. The  simulation trajectories were previously generated by our group in multiple past works\cite{Mukherjee_2019_32bead,Ahalawat_2018_gb1,Mukherjee_2020_trpcage}.  For 32-bead-in-a-spring charge-neutral polymer (with alternatively patterned positive and negative charges on the beads), 200 MD simulation trajectories, each 30 ns long, modelled in aqueous media were used as inputs. The model and simulation details of the polymer system have been reported in a previous investigation from our research group\cite{Mukherjee_2019_32bead} .  Likewise, For GB1 $\beta$-hairpin, 200  previously performed MD trajectories\cite{Ahalawat_2018_gb1} , each 100 ns long, served as the input. Finally 200  MD trajectories of conformational fluctuation of Trpcage miniprotein, each 100 ns long in neat water as well as in 4 M Trimethyl amine N-oxide (TMAO) act as the resource for CV discovery in this system. The method and model details used for generating all MD trajectories of Trp-cage have been reported in a recent article by Mukherjee and Mondal\cite{Mukherjee_2020_trpcage}.   
   
   \textit{Methods:}

\begin{figure*}[htp]\centering
\includegraphics[height=7.5in,width=7.5in,keepaspectratio]{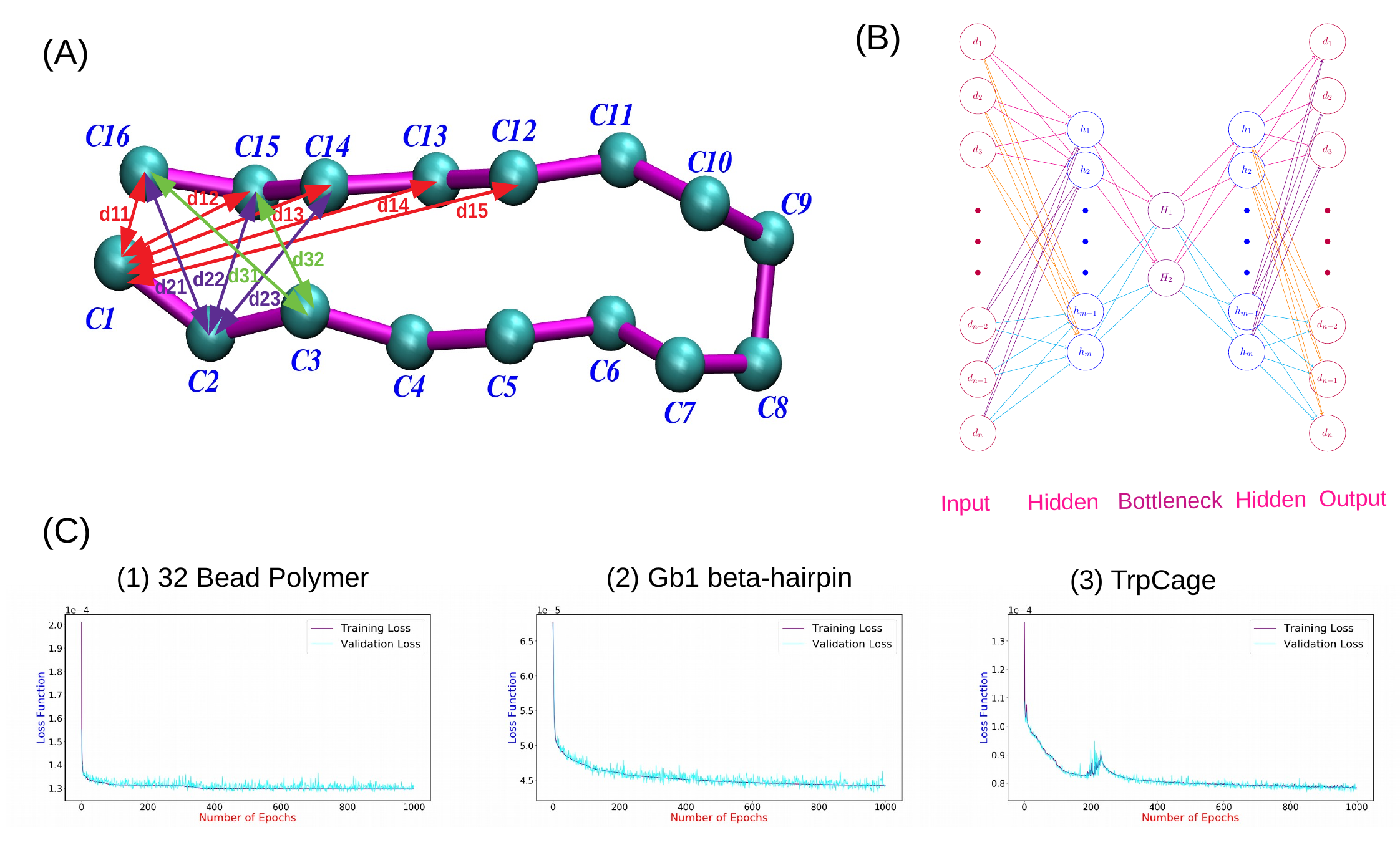}
\caption{ (A)Input CVs: A set of inter-residue $C\alpha$-distances   (B) Schematics of auto-encoder scheme employed in the work. The set of inter-residue  $C\alpha$-distances described in figure A is used here as the input and the dimensions are iteratively reduced with a focus on minimising the loss function. (C) The profile of loss function as obtained on course to an optimized auto-encoding of input CVs for three systems investigated here.   }
\label{cv}
\end{figure*}

        The current work focuses on implementing the frame-work of autoencoder based artificial neural network in reducing the large number of dimensions of a set of biomacromolecules (see Figure~\ref{snap}) into a low number of optimal CVs.      In stead of limiting ourselves to traditionally preconceived CVs for protein folding (such as Rg,RMSD,Number of contacts etc), here we opted for more generic features as input CVs for our neural-network auto-encoders. More specifically, the large combinatorial set of inter-residue distances for each systems of interest served as the input CVs of the autoencoder applied in the current article (See Figure~\ref{cv}A for illustration of the input CVs).   
        
        Autoencoder is a self-supervised machine learning technique where the neural network would strive to generate the final output which is same as the supplied input. Figure~\ref{cv}B schematically depicts the architecture of the autoencoder employed in the current investigation. By construction, these autoencoders would have an input layer, an output layer and one or multiple hidden layers in between the input and output layers with a bottleneck in the middle.  The layer structure is usually symmetric, with the first half including the bottleneck referred as the \textit{encoder}, while the second half is called a \textit{decoder}. The input layer and output layer have same number of nodes while the intermediate hidden layers in encoder and decoder part often have fewer number of nodes, thereby forcing the network to reduce the dimensions and consequently to learn a lower-dimensional, compressed representation in the latent space from the present data-set with a minimal regression error. For each pair of nodes there is an associated weight ($w_{ij}$). The linear combination of the inputs with appropriate weights after a bias addition and followed by some activation function operation will result in the value of the output on next node. This activation function can be defined by:
      	     \begin{equation}
      	     	x_{i}^{l+1} = f\LARGE(\Sigma_{j=1}^{N} w_{ij}^{l} x_{j}^{l} + b_{i}^{l+1})
      	     \end{equation}
      	     ,where $x_{i}^{l+1}$ is $i$-th node at $(l+1)$-th layer, $x_{j}^{l}$ is $j$-th node at $l$-th layer, $N$ is the number of nodes in $l$-th layer, $w_{ij}^{l}$ is the weight corresponding to the $(i,j)$-th pairs of nodes at $l$-th layer, $b_{i}^{l+1}$ is the bias at the $i$-th node at $(l+1)$-th layer.

  The $C_{\alpha}$-$C_{\alpha}$ inter-residue distances have been used as input and output data in this work for the feedforward densely connected neural network. The specific number of dimensions used in each layer has been mentioned in respective example of biomacromolecules in Results section.  0.1 fraction of the data has been taken as validation data set while the rest was kept as training data set. We have used symmetrical dense neural network for our work.  Activation functions are crucial in order to introduce the non-linearity in the network, hence facilitating the feature of non-linear fitting into the model on the data-set. In the current work, the activation functions were used as $tanh$ and $linear$ alternatively from input to output layers.  Here the weights and biases as described above are the parameters which have to be optimised in order to minimise a loss function. In our work the loss function or regression error is used as mean squared error(MSE) and it is defined as:
   $ MSE = \Large \Sigma_{i=1}^{n}(y_{i}^{true}-y_{i}^{predicted})^2 /n $ 
    ,where $n$ being the total number of samples. For 32-bead polymer, gb-1 and Trp-cage systems 6000200 ; 2000200 and 2000200 samples were used respectively.  Our model has been implemented using tensorflow \cite{tensorflow} backend with Keras \cite{Keras}  Python Library for Neural Networks. The optimizer used here is Adam\cite{ADAM} with learning rate $\eta =0.001$ and other parameters used were $\beta_1=0.9$,$\beta_2=0.999$,$\epsilon=1e-07$ with batch size as 100 to train our model. The weights were initialized using glorot uniform method\cite{pmlr-v9-glorot10a} and biases were initialised as zeros. The latent space is considered our desired CVs, as obtained from auto-encoder. The structure and the number of the layers in encoder and decoder parts of the neural networks have been described in Results section in context to each of the systems investigated in the work.

      In the current work, we have compared the autoencoder based dimensional reduction approaches with two popular linear dimension reduction approaches, namely PCA and TICA. In both cases, the same set of combinatorial inter-residue distances served as input feature space. Below, for the sake of completeness,  we provide a brief discussion of PCA and TICA.
      
 \textit{PCA:}
 Principal Component Analysis(PCA) is a widely used \cite{PCA_ref1} feature extraction tool leading to dimensional reduction technique. Here the goal is to derive those dimensions which will maximize the variance along it by linearly combining the input feature vectors. This procedure of maximizing the variance is equivalent to obtain the eigen components of the covariance matrix from the input data set. Eigen vectors obtained after diagonalization of the covariance matrix leads to the coefficients of that linear combination. The corresponding eigen values will depict the explained variance along that projected mode. Mathematically, PCA  would maximize the variance from a given input data-set: \newline \\\
 $$ Var(x) = \frac{\Sigma_{i=1}^{N}(x_{i}-\bar{x})^2 }{N} $$ \\\
 ,where Var(x) is the variance along x-dimension, $x_{i}$ is the i-th value of a time-series data. Index i describes the time-series index and varies from 1 to N as there are N time points present here for example and $\bar{x}$ is the average value of the x-dimension data,
 $\bar{x} = \Sigma_{i=1}^{N} x_{i}/N $. \newline
 PCA then maximizes  $Var(x)$.
 If there are $m$ such input dimensions, PCA would determine $m$ linearly combined projected variables. The total sum of variances would be distributed among them in a way such that the first principal component(PC1) will have maximum variance. The retained variance again would be distributed in this way, making the second principal component (PC2) to have next maximum variance from the rest of the total variance and so on. 
 
  \textit{TICA:} 
  The Time-lagged Independent Component Analysis (TICA) is another linear dimension reduction technique which has been applied successfully for bio-molecular simulation \cite{TICA_ref1,TICA_ref2}. Originally  used in signal processing\cite{Molgedey_1994_tica},TICA is mostly similar to the previously explained PCA technique. The main difference is that instead of maximizing the variance, TICA will maximize a time-lagged autocorrelation function. TICA transforms and projects the input data along those dimensions for which the autocorrelation function is the maximum. Mathematically speaking , TICA  will maximize the time-lagged autocorrelation function as depicted below: \newline
{
$$ Corr(x) =  \frac{  [ \Sigma_{i=1}^{N-\tau}(x_{i}-\bar{x})(x_{i+\tau}-\bar{x}) ] / (N-\tau) }{ \sigma_{x}^{2} } $$
}
,where $Corr(x)$ is the time-lagged kinetic variance along x-dimension, $x_{i}$ is the i-th value and $x_{i+\tau}$ is the $(i+\tau)$-th value of a time-series data, $\tau$ is the lag interval. Index i varies from 1 to $N-\tau$ as there are $N-\tau$ time-series data points after considering a lag-interval of $\tau$. $\bar{x}$ is the average value of the x-dimension data,
$\bar{x} = \Sigma_{i=1}^{N} x_{i}/N $ . Again $\sigma_{x}$ is the standard deviation along x defined by: $ \sigma_{x} = \sqrt{ \frac{\Sigma_{i=1}^{N}(x_{i}-\bar{x})^2 }{N}} $. \newline
Here TICA would maximise this time-lagged correlation function $Corr(x)$.
If the data-set consists of $r$ such input dimensions, then TICA will find out $r$ linear combinations among them. The total sum of kinetic-variances will be distributed in such a way that the TIC-1 component will get maximum kinetic-variance and consequently will become kinetically most sluggish. The retained kinetic-variance again will be distributed in this way so the TIC-2 will become next slowest component and so on.

Finally, we use the variational approach for Markov processes (VAMP-2) score\cite{vamp_ref1,vamp_ref2} to compare the optimality of the auto-encoder-derived latent space CVs relative to PCA and TICA. 

\textit{VAMP-2 score:} VAMP is based on Canonical Correlation Analysis(CCA)\cite{CCA} method for time-series data and also known as Time-lagged Canonical Correlation Analysis(TCCA). Here we will look for the cross correlation function between a pair of of variables, say for example (x,y). CCA will maximize the cross-corelation function for this pair. Analytically, \newline
	$$ Corr(x,y) =  \frac{  [ \Sigma_{i=1}^{N}(x_{i}-\bar{x})(y_{i}-\bar{y}) ] / N }{ \sigma_{x} \sigma_{y} } $$
	,where $Corr(x,y)$ is the cross correlation function between (x,y). $x_{i}$ is the i-th value for x-variable and $y_{i}$ is the i-th value of the y-variable. $\bar{x}$ and $\bar{y}$ are the average values of the x and y respectively. Again $\sigma_{x}$ and $\sigma_{y}$ are the standard deviations along x and y respectively. \newline
	Here CCA will maximize this cross-correlation function $Corr(x,y)$. Now, for time-lagged CCA, we will typically divide our data-set into two halves. For first half, we will have our data as x and index i goes from 1 to $(N-\tau)$. For the last half, we will have our data as y and here index i goes from $(\tau+1)$ to $N$. So, eventually it would involve a time-lag interval of $\tau$, hence the name as time-lagged CCA(TCCA). The VAMP-2 score will be the sum of the diagonal elements of that time-lagged cross-correlation matrix(after diagonalization) raised on a power of 2(squared). The higher the value of this score implies the better it is conserving the slowest modes and hence giving rise to optimality of the collective variable.

       Apart from the regular projection of the trajectories along the chosen CVs, in some cases, the MD simulation trajectories were discretised and clustered along the derived feature spaces and a Markov state model (MSM)\cite{Prinz_2011_MSM_ref1,Bowman_2009_MSM_ref2}  is built to analyze the free energetics and kinetics of metastable conformational ensembles. We have described the MSM protocols in the Results section.

\section{Results and Discussion}

 \subsection{32-Bead Polymer:}

  \begin{figure}[H]
  	\includegraphics[clip,width=1.0\columnwidth,keepaspectratio]{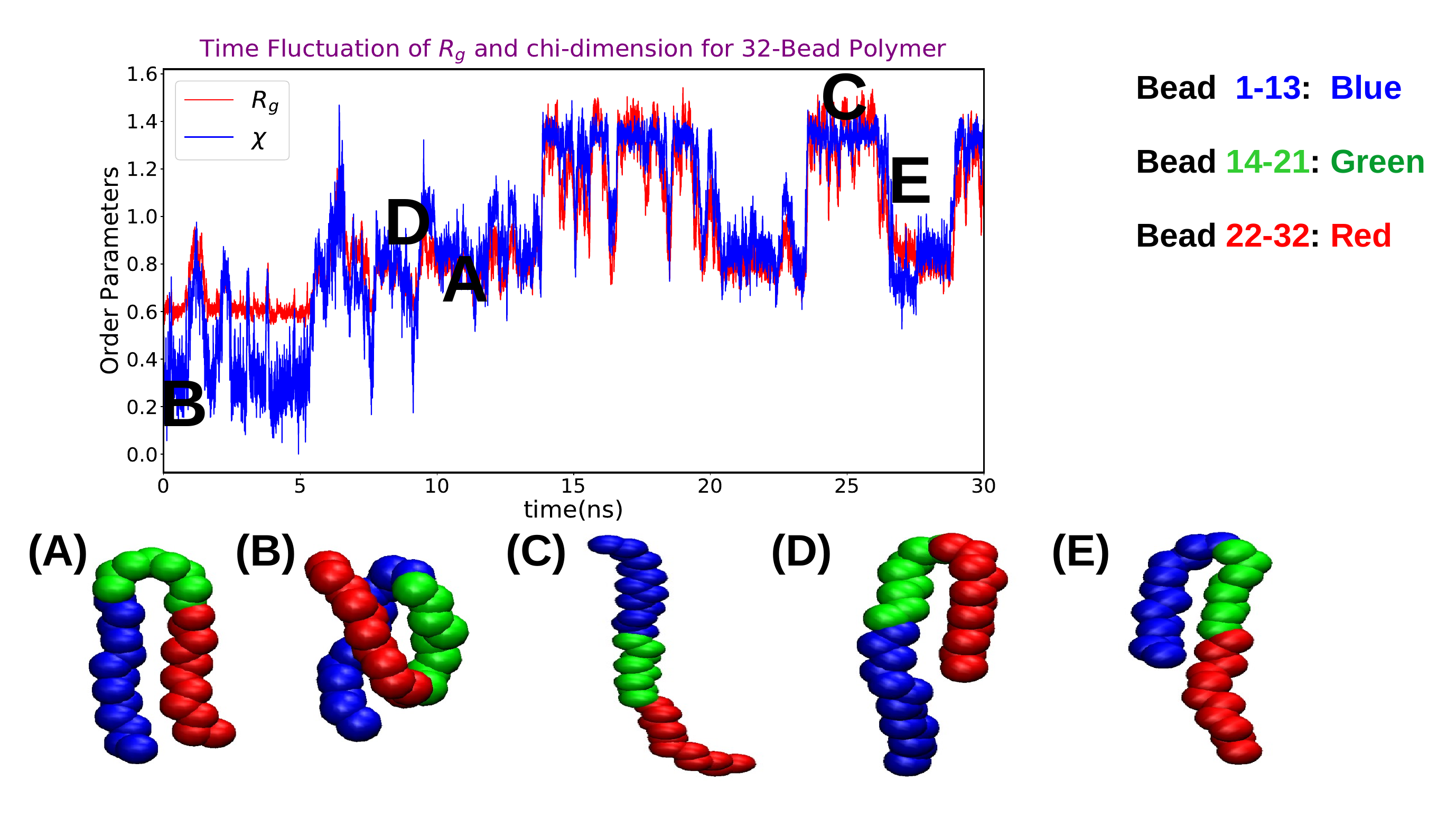}
  	\caption{ The proposed Autoencoder Scheme captures time-fluctuation of the different conformations for 32-bead polymer system . A,B,C,D,E are the diffterent conformations of the polymer.  }
  	\label{poly}
  \end{figure}
  
   We first assess the ability of the proposed input CVs and the autoencoder-derived latent-space dimension in capturing the essential conformational dynamics of  a prototypical model polymer ( see Figure~\ref{snap}A). This 32-bead-in-a-string polymer and its variants have remained an attractive system of interest in multiple precedent applications.\cite{Mondal_2013,Mukherjee_2019_32bead} Despite its apparent simplicity, this polymer is known for displaying interesting dynamical conformational transitions. The polymer of interest in this work has 16 alternatively positive and negative charged beads, rendering the complete system charge-neutral. The presence of charges allow for increased conformational fluctuation of this polymer in aqueous media. The aggregated 6 microsecond of MD simulation trajectories (200 independent trajectories, each 30 ns long) serve as the source. In any case, even this system with  $\Comb{32}{2} = 496 $ number of pair-wise distance brings out a large dimensionality, which is well beyond what is trivially graspable and is still fairly simple model. To apply our method for broader scope of application and in order to use this method in a more tractable way, we started with $28$ number of pair-wise distances as input features or CVs,  instead of starting with all the $ 496 $ pair-wise distances for the autoencoder. In particular, we have chosen the beads starting from the first bead in an arithmetic progression with an interval of 4, hence total number of effective beads we have obtained to be 7 among the 32 beads. So, we have got $\Comb{7}{2} = 56/2 = 28 $ dimensional pair-wise distances.    We find that an autoencoder of $5$ layers having nodes 28,12,4 for the encoder part, with a  latent space combination of  $4$ CVs gradually reduces the loss-function to a plateau  (see Figure~\ref{cv} C). The decoder used in our work had same architectures of layers as in encoder, except in reverse fashion.

  In Figure~\ref{poly}, a representative MD trajectory of the 32-bead polymer is projected on latent space CV of the model polymer, as derived by the autoencoder upon minimisation of loss function or the regression error. The time profile of the derived latent space CV of the model polymer (Figure~\ref{poly}) indicates multiple dynamical transitions. In particular, an overlay of the time-profile of complex latent space dimension with that of a traditional CV, namely, radius of gyration ($R_g$), suggests that the derived CV recapitulates the salient trend of conformational dynamics quite well. The time profile of latent space projection captures a series of conformational ensemble of the polymers, namely collapsed, extended, hairpin and partial extended states. The faithful reproduction of conformational landscape of this polymer by the autoencoder derived CVs and its similarity with a popular knowledge-based CV like $R_g$ for polymer, indicates that the derived latent-space CV shows early promises of discovering new conformation in relatively complex biomacromolecules.

     \subsection{GB1 $\beta$-hairpin:}
     
        The demonstrated ability of the frame-work in spatial separation of conformational landscape of model polymer prompted us to apply the protocol to explore the  dynamical interplay of a small but complex polypeptide, namely 16-residue GB1 $\beta$-hairpin. This particular polypeptide has been the subject of numerous experimental \cite{Blanco_1994_Nature_structural_mol_biology,Soranno_2018_jpcb_recent} and computational\cite{Juraszek2009,Best2011,Andrec2005,DanielSWeinstock2007} investigations due to the diversity of metastable conformations it demonstrates. Our past investigation has also suggested that a single CV is not sufficient for capturing the underlying conformational complexity of this system.  Rather an optimized combination of a set of curated CVs is a necessity.\cite{Ahalawat_2018_gb1,Prakashchand_2020_gnap}
 
 \begin{figure}[H]
 	\includegraphics[clip,width=1.0\columnwidth,keepaspectratio]{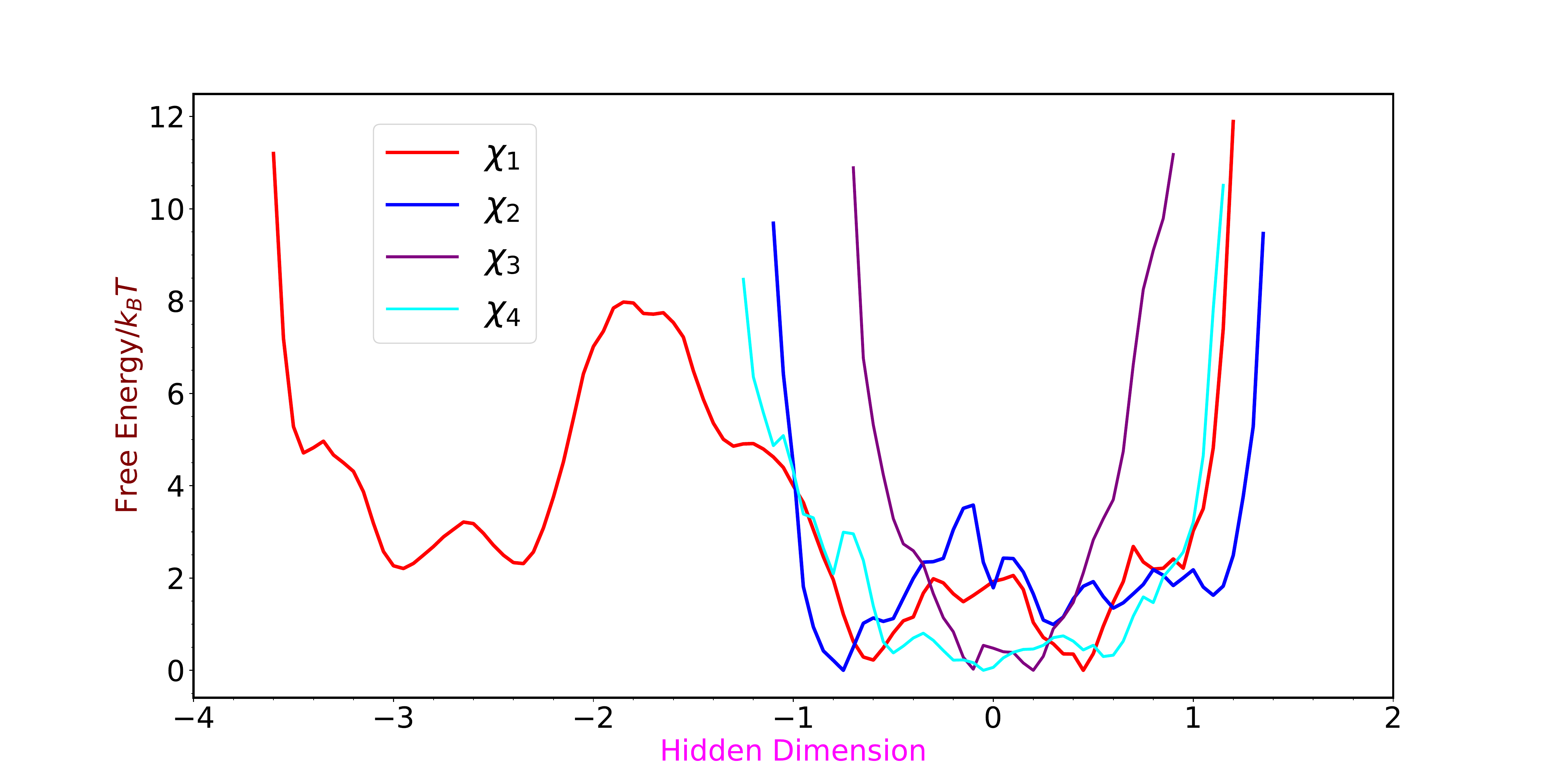}
 	\caption{One Dimensional Free Energy Distributions for Hidden data for GB1.}
 	\label{1d}
 \end{figure}

\begin{figure*}[htp]\centering
  	\includegraphics[height=7.5in,width=7.5in,keepaspectratio]{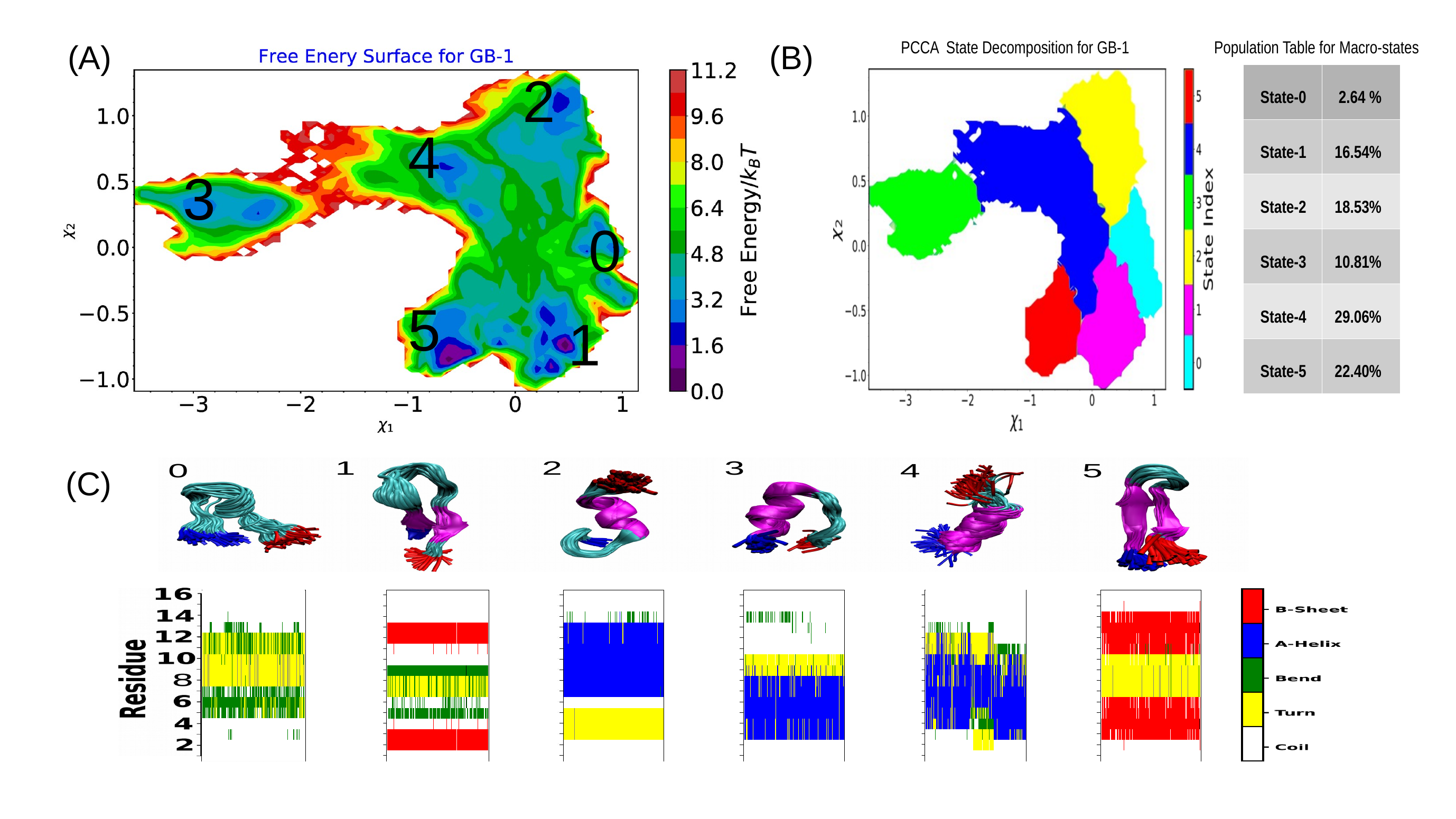}
  	\caption{ (A) Free energy surface of conformational landscape of GB1 $\beta$-hairpin along two of the four latent CVs \newline (B) The PCCA State Decomposition Diagram along with Population Table  (C) The representative conformations corresponding to the free energy basins. For each basin, the secondary structure of all conformations are also characterised via DSSP scores.   }
  	\label{free}
\end{figure*}

In this context, we wanted to investigate  if an auto encoder frame-work with a large set of inter-residue distances as input features, can spatially and temporally resolve the metastable ensembles of GB1 $\beta$-hairpin. Towards this end, we employed all pair-wise distances between C-$\alpha$ atoms as input dimensions in the auto-encoder, giving rise to $\Comb{16}{2} = 120 $ input dimensions for a 16-residue poly-peptide. We have invoked these large number of input dimensions in the auto encoder and slowly quenched these to a very less number of hidden dimensions. For this purpose, 72, 36,12 number of  nodes were used in the first, second and third hidden layers respectively. Finally 4 nodes were used in the fourth and most compressed hidden layer, constituting the latent space bottleneck in the encoder part. The data analysed was on the projected  data along this few number of latent dimensions, hence satisfying the goal of extracting features from big-data sets via dimensional reduction. The autoencoder is  symmetric around the bottleneck, the decoder part also adds same number of hidden layer as the encoder part in reverse order  ( i.e. 12,36,72 number of hidden nodes), followed by 120 dimensional nodes as output.

  The comparison of one-dimensional projection of free energy profiles along each of the four latent-space CVs ( Figure~\ref{1d}) suggested that combination of two CVs ($\chi_1$ and $\chi_2$) will be sufficient for effective projection of MD trajectories.  Accordingly, we project the free energy surface of GB1 along $\chi_1$ and $\chi_2$. Figure~\ref{free} A) depicts the two-dimensional free energy surface (FES) along these two auto-encoded latent spaces. We find that the projection of FES along the chosen latent-space dimension spatially resolved multiple basins. Visual inspection of the conformational ensembles, extracted from each annotated basin in the FES, mutually separates out the diverse conformational ensembles of GB1 $\beta$-hairpin. Specifically, the free energy decomposition recovered  metastable conformations ranging from $\beta$-sheets, partially folded helices to unfolded coils. A secondary structure analysis , based on DSSP algorithm\cite{Kabsch_1983_DSSP} of conformational ensembles (Figure~\ref{free} C) further ascertained that these are distinct conformations having little overlap in the features of their secondary structures. The secondary structures of each of the conformations remain unmixed in the free energy surface projected along the latent space.

    \begin{figure*}[htp]\centering
    	\includegraphics[height=7.5in,width=7.5in,keepaspectratio]{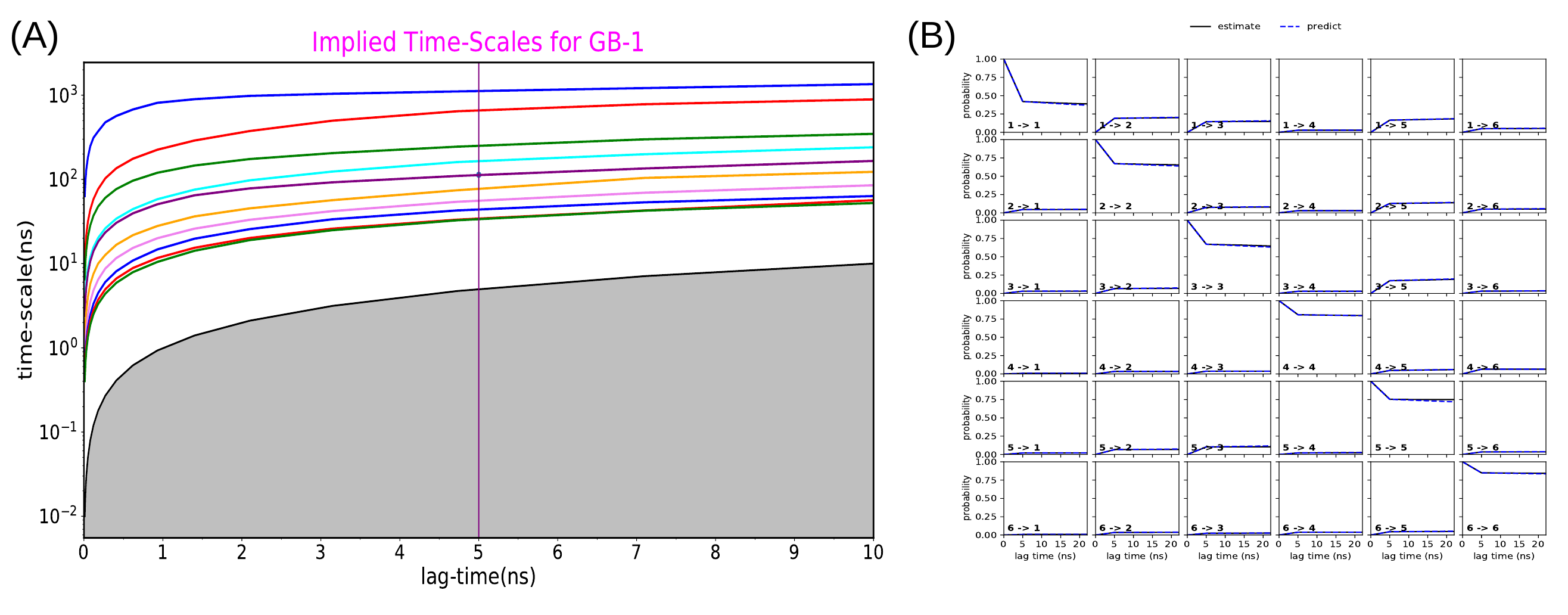}
    	\caption{ Using Encoded Data: (A) Implied Time-scale plot (B) Chapman-Kolmogorov (CK) test.}
    	\label{its}
    \end{figure*}

    The ability of these neural-network derived CVs in spatially resolving the FES encouraged us to optimally tessellate the metastable states of GB1 $\beta$-hairpin along the projected dimension. Towards this end, we discretised all simulation trajectories of GB1 $\beta$-hairpin  using the two latent space CVs into 500 micro states via k-means clustering\cite{kmeans} and built a MSM. The implied time scale (ITS) underlying the MSM built in the reduced conformational sub-space was found to readily converge to a plateau within a short lag-time, representing a clear time-scale separation among six metastable macrostates (Figure~\ref{its}A) . Additionally, as shown in Figure~\ref{its}B) the six macrostate model also passed Chapman-Kolmogorov (CK) test seamlessly. These analysis indicated that the model based on the autoencoder-derived CVs is robust as far as attaining the Markovianity is concerned. Our Markov model has been built with 5 nano-second of lag-time. A PCCA\cite{pcca_2006} based coarse-graining of $500$ microstates  into six macrostates tile them on distinct location of the FES.

 \begin{figure} 
  	\includegraphics[clip,width=1.0\columnwidth,keepaspectratio]{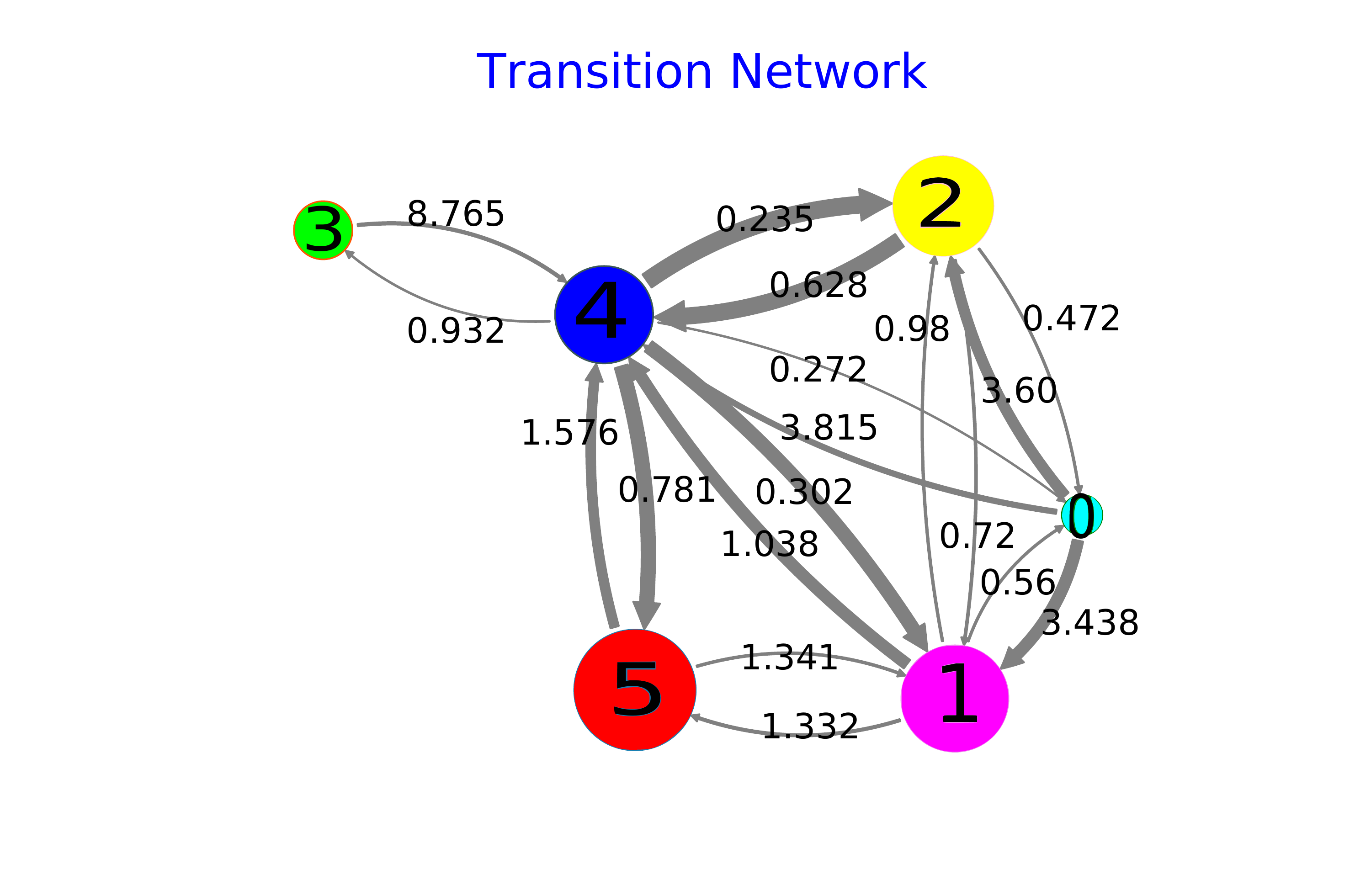}
    \caption{ Transition Network for GB-1 with the MFPT values written in $\mu$s. Refer to Figure~\ref{free} for index of macro states.}
    \label{network}
 \end{figure}  
    
    These macro-states are mutually distinct on secondary structural basis. We confirm this by rendering the representative snapshots of each of the macro states and via quantifying the secondary structure of each of the macrostates via DSSP algorithm. Our analysis and the visual inspection indicate that although the crystal structure of GB1 is suggestive of $\beta$-hairpin, the conformational ensemble of GB1 peptide segment is highly heterogeneous. However interestingly, the adopted latent space CVs are able to distinctly identify each individual macrostates.  The resolved FES and its PCCA decomposition show that along with perfect $\beta$-sheet conformation (Macrostate 5), a large array of conformations, namely, distorted $\beta$-sheet (Macrostate 1), N-terminal $\alpha$-helix (Macro-state 3), partial $\alpha$-helix (Macrostate 4), C-terminal $\alpha$-helix(Macrostate 2) and random coil(Macrostate 0) (See Figure~\ref{free} A-C) also coexist.  Folded GB-1 exists in perfect $\beta$-sheet conformation($22.40 \%$). Additionally, the state-space decomposition suggests GB1 can adopt another $\beta$-hairpin like conformation, distorted $\beta$-sheet($16.54\%$) in the ensemble. The full population of $\beta$-hairpin like conformations is combined to be $38.94\%$. These are in good agreement with the experimentally measured population of $\beta$-hairpin conformations\cite{Blanco_1994_Nature_structural_mol_biology,Soranno_2018_jpcb_recent}. Previously, computer simulations based on structure based clustering~\cite{Best2011} and the sketch-map analysis~\cite{Ardevol_2015_sketch_map} of simulated data had suggested that this polypeptide could have more than four different types of metastable conformations i.e, unfolded, collapsed, helical, and $\beta$-sheet structures. But none of the previously proposed CVs, on its own, has been able to dissect all of these conformationally distinct states. In this regard, the current framework, via combining auto encoder based deep learning approaches with Markov state model, is found to identify and distinguish the native and native-like conformations of a dynamically fluctuating polypeptide like GB1 $\beta$-hairpin. In addition the projection can potentially capture the subtle non-native conformational fluctuations for this small peptide, thereby manifesting the efficacy of the identified collective variables. The spatial resolution of the FES, derived in the current investigation, is found to be significantly higher than that obtained via sketch map for the same system\cite{Ardevol_2015_sketch_map}.

  \begin{figure*}\centering
     \includegraphics[height=7.5in,width=7.5in,keepaspectratio]{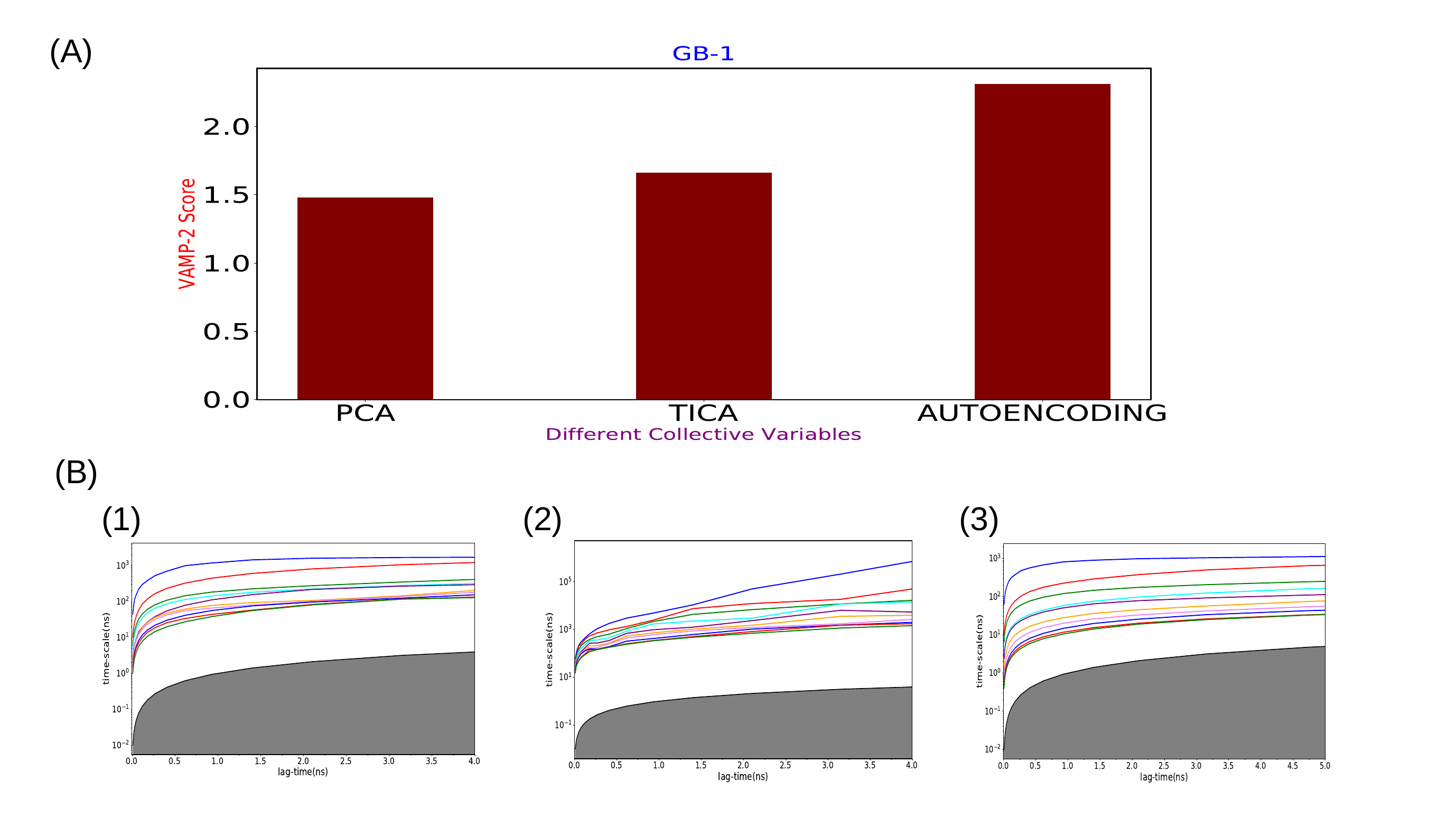}
 	 \caption{ Comparison of auto-encoder-derived CVs with PCA and TICA (A) via VAMP analysis  (B) Using ITS Plots (1) ITS for PCA projected data (2) ITS for TICA projected data (3) ITS for Encoded data }
  	 \label{vamp}
  \end{figure*}

  	 After coarse-graining of the micro-states to 6 different macro-states we have calculated the mean first passage time(mfpt) of transitions between different macro-states. Random coil has been taken as unfolded state and perfect $\beta$-sheet as native folded conformation. The folding time (unfolded to folded) has turned out to be 4.40 $\mu$-second and the unfolding time (folded to unfolded) has turned out to be 1.48 $\mu$-second. Both these values are in good agreement with the experimentally measured rates or transition times\cite{Mu_oz_1997_gb1_kinetics} for GB1 $\beta$-hairpin. Then we have performed transition path thoery(TPT)\cite{Noe_2009_TPT_PNAS} analysis on this $\beta$-peptide system to resolve the kinetic pathways for this folding-transtions from unfolded to folded state. We have found that there are 4-major pathways for this folding to occur (see Figure~\ref{network}).  Firstly, the transition can happen through Random Coil, C-terminal $\alpha$-helix, partial $\alpha$-helix, perfect $\beta$-sheet  [$0 \rightarrow 2 \rightarrow 4 \rightarrow 5$] with a percentage of pathways as 31.5\%. Secondly, 22.7\% of pathways are following a pathway with a sequence of Random Coil, distorted $\beta$-sheet, partial $\alpha$-helix, perfect $\beta$-sheet meaning [$0 \rightarrow 1 \rightarrow 4 \rightarrow 5$]. Thirdly, 20.0 \% of total pathways are following Random Coil, distorted $\beta$-sheet, perfect $\beta$-sheet  [$0 \rightarrow 1 \rightarrow 5$]. Finally, 19.3 \% of total pathways are following through Random Coil, partial $\alpha$-helix, perfect $\beta$-sheet meaning [$0 \rightarrow 4 \rightarrow 5$]. These 4 pathways are contributing altogether 93.5 \% of total folding pathways.

      How do the latent space CVs derived using the neural-network approach rank in comparison with the popular independent component analysis techniques? Towards this end,  we individually perform PCA\cite{PCA_ref1} and TICA \cite{TICA_ref1} on the original input CVs (set of pair-wise inter-residue distances) of GB1 $\beta$-hairpin.  Contrary to neural-network based approaches PCA and TICA are linear dimensional reduction techniques. However, PCA focusses on maximising the variance while TICA strives to maximize the time-correlation of the projected dimension. For a robust comparison among PCA, TICA and auto-encoder derived optimised CVs on an equal footing, we employ variational approach for Markov processes (VAMP-2) score, as introduced by Noe and coworkers\cite{vamp_ref1,vamp_ref2}. VAMP-2 score enables one to cross-validate multiple candidate CVs for their optimal representation and have been instrumental in making approbate choice of CVs in describing biomolecular conformation\cite{bhu}.  For a quantitative measure of relative optimality of the CVs, we analyze the VAMP-2 score on the  CVs derived from each of the three dimension-reduction techniques (PCA,TICA and autoencoder). As shown in Figure~\ref{vamp} A, the VAMP-2 score is clearly the highest for the latent space CVs derived from neural-network based auto-encoder, compared to TICA and PCA. Together this suggests that the quality of auto-encoder derived latent space CVs would be superior over TICA or PCA-derived CVs (applied on same input features) for efficient exploration of conformational landscape of GB1
       $\beta$-hairpin. Again the comparison of ITS plots has been shown in Figure~\ref{vamp} B for the PCA, TICA and Autoencoder-projected data to show the greater ease of having clear implied time scale separation attaining plateau for the Encoded-data.

 \subsection{TrpCage Mini-protein:}
   
  Finally, as an application of auto-encoder derived CVs, we investigate if these CVs can assist in elucidating the role of osmolyte in tuning the conformational landscape of the  proteins. Here-in, we apply the auto encoder-based optimized CVs to assess the effect of popular osmolyte Trimethyl amine N-oxide (TMAO) on the conformational landscape of  mini protein Trp-cage. The previously performed\cite{Mukherjee_2020_trpcage} swarm of MD simulation trajectories of Trp-cage miniprotein in 0 M (i.e. neat water) and 4 M aqueous TMAO formed the basis of our current investigation. Consistent with protocols deployed in the other two systems, we use pair-wise inter-residue distances of Trp-cage mini protein as the input CVs in our auto-encoder frame-work and employed a similar auto encoder schemes as in GB1 $\beta$-hairpin for deriving the latent space variables.

  \begin{figure*}[htp]\centering
   	\includegraphics[height=7.0in,width=7.0in,keepaspectratio]{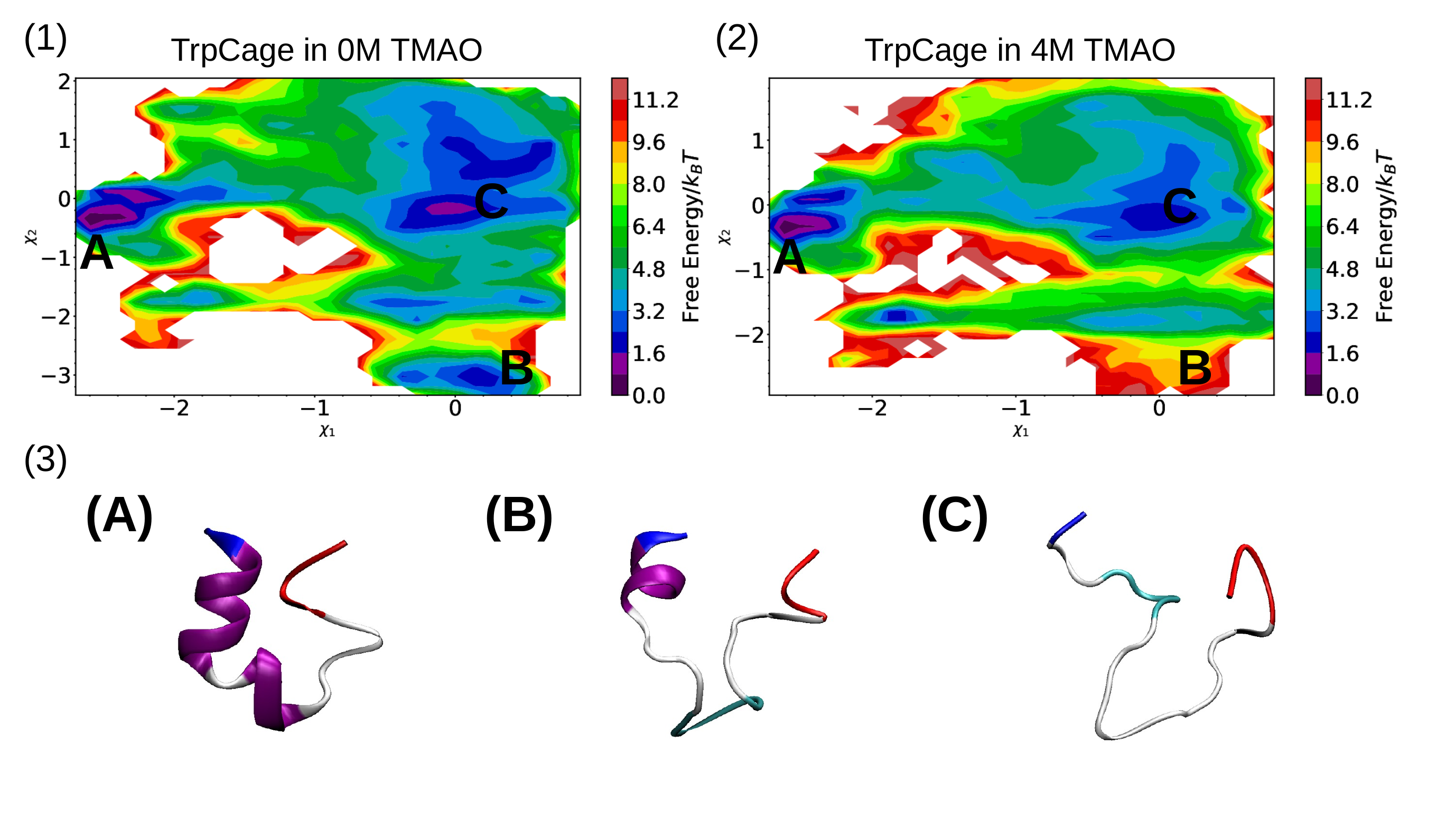}
   	\caption{(1) Projection of conformational landscape along latent CVs of Trp-Cage in neat water. (2) The projection along same latent CVs in 4 M TMAO. (3) The Representative Snap-shots for TrpCage.  }
  	\label{trp}
  \end{figure*}
  
     Figure~\ref{trp} (1) represents the two-dimensional projection of  conformational space of Trp-cage in neat water along the latent space $\chi_1$ and $\chi_2$. Three free energy basins corresponding to the native fold, partially folded and unfolded states of the Trp-cage is evident in the FES in neat water. Interestingly, as shown in Figure~\ref{trp} (2), the projection of MD trajectories of Trp-cage in 4 M aqueous TMAO solution along the same latent space $\chi_1$ and $\chi_2$ (which was obtained using the trajectories in neat water) clearly indicates that the presence of TMAO free energetically destabilises the basin corresponding to the partially folded and unfolded state, thereby tilting the free energy more favourably towards the native folded state of the Trp-Cage.  Together, this analysis dictates that the CVs discovered via the autoencoder based neural network is able to distinguish the conformational landscape in different media and pinpoint the key factors.
      
     The present scheme recovers the fully folded $\alpha$-helix, unfolded random coils and partially folded $\alpha$-helix conformations as three major macro-states following the principle of Implied Time Scale plots.  For the folded  states these structures have good resemblance  with the previously identified structures for the same system\cite{Chen_2018_cv_discovery_in_enhanced_sampling}, albeit using enhanced sampling methods. However, the extent of diversity in the  metastable states discovered , especially the coexistence of unfolded random coils and partially folded $\alpha$-helix  observed in the current work is significantly more than that in  previous work.\cite{Chen_2018_cv_discovery_in_enhanced_sampling}  The extent of spatial separation of the conformations, derived in the current setup is also significantly superior than previously adopted encoder map\cite{Lemke_2019_Encoder_Map}.

\section{Conclusions}

In summary, the current work employs a generic input CVs, namely an exhaustive sets of inter-residue distance of a biomacromolecule to derive an optimal low-dimensional embedded CVs within the frame-work of auto associative deep learning neural network. By striving to encode N-dimensional input features as a d-dimensional representation (d $<<$ N), the protocol passes the information through
the bottleneck and reconstructs the original signal again in the
decoder. The resultant latent space CVs are shown to resolve the crucial conformations of three hierarchically complex systems of biomacromolecules. In combination with MSM, the latent space CVs are able to distinctly identify the key macro-states and avoid any mixing. The approach is able to sense the effect of solvent and cosolute in tuning the conformational  landscape of the protein. Finally, the superiority of the auto-encoded latent space over linear combination based techniques (namely PCA and TICA) is rigorously demonstrated. 

While  extensions of linear dimensional reduction techniques ,such as  Kernel PCA and kernel TICA\cite{Schwantes_2015_kernel_methods}, present a means to alleviate this problem by applying a known nonlinear transformation of the atomic coordinates prior to dimensionality
reduction, the specification of appropriate kernels can be almost as tedious job as guessing the CVs themselves.  In this regard, the frame-work presented in the current article provides a practical and efficient avenue for CV discovery and identification of kinetically relevant metastable conformations. The frame-work of auto encoding based approach, similar to what has been described here has recently been found to be quite effective for CV discovery and for generating synthetic trajectories via simulating the latent space\cite{Sidky_2020_latent_space_simulator}.  A related approach in this direction has enabled on-the-fly CV discovery and accelerated  energy landscape exploration via combining umbrella sampling with autoencoder-based dimensional reduction\cite{Chen_2018_Ferguson_Computational_Chemistry}. On the contrary, in this work, we have  avoided the introduction of enhanced simulation approaches, which may run the risk of introducing biases on CV discovery.  We demonstrate that the usage of unbiased MD simulation trajectories for identifying appropriate CVs within the frame-work of auto encoder and its effective combination with MSM leads to clearer separation of the conformational landscape. The higher spatial and temporal separation of the conformations is particularly evident for GB1 $\beta$-hairpin when compared to that obtained from sketch map\cite{Ardevol_2015_sketch_map} .  Apart from the elucidation of static conformational landscapes via discovery of optimal CVs, emerging directions have seen extension and proposals of suitable framework for capturing the underlying dynamics. Development of relevant deep-learning based techniques such as  time-lagged autoencoders\cite{Wehmeyer_2018_Noe_TAE,Chen_2019_time_lagged_encoder}, Variational dynamical encoders\cite{Hern_ndez_2018_variation_encoding}, RAVE\cite{Ribeiro_2018_rave}  have paved the way for prediction of future dynamics and reconstruction of  dynamical trajectories for macromolecules. Future directions will aim to extend along this line.

\section*{Acknowledgments}
This work was supported by computing resources obtained from shared facility of TIFR Centre for Interdisciplinary Sciences, India. We acknowledge support of the Department of Atomic Energy, Government 
of India, under Project Identification No. RTI 4007. JM acknowledges Ramanujan Fellowship and Core Research grants provided by the Department of Science and Technology (DST) of India (CRG/2019/001219). SB thanks to TIFR for providing all support. SB also thanks Dr. Navjeet Ahalawat for  useful discussions on related topics.

\section*{References}

\bibliography{MS}

\end{document}